\documentstyle[prl,aps,floats,epsf]{revtex}
 
\begin{document}
 
\draft                    
 
\wideabs{
 
 
\title{
         \vskip -0.5cm
         \hfill\hfil{\rm\normalsize Printed on \today}\\
         Unusually High Thermal Conductivity of Carbon Nanotubes}
 
\author{ Savas Berber,
         Young-Kyun Kwon,
         and David Tom\'anek}
 
\address{Department of Physics and Astronomy, and
         Center for Fundamental Materials Research, \\
         Michigan State University,
         East Lansing, Michigan 48824-1116}
 
\date{Received 23 February 2000 }
\maketitle
 
 
\begin{abstract}
Combining equilibrium and non-equilibrium molecular dynamics
simulations with accurate carbon potentials, we determine the thermal
conductivity $\lambda$ of carbon nanotubes and its dependence on
temperature. Our results suggest an unusually high value
${\lambda}{\approx}6,600$~W/m$\cdot$K for an isolated $(10,10)$
nanotube at room temperature, comparable to the thermal conductivity
of a hypothetical isolated graphene monolayer or diamond. Our results
suggest that these high values of $\lambda$ are associated with the
large phonon mean free paths in these systems; substantially lower
values are predicted and observed for the basal plane of bulk
graphite.
\end{abstract}
 
 
\pacs{
61.48.+c,
%
66.70.+f,
%
63.22.+m,
%
%
68.70.+w
%
 }
 
 
}
 
\narrowtext
With the continually decreasing size of electronic and micromechanical
devices, there is an increasing interest in materials that conduct
heat efficiently, thus preventing structural damage. The stiff $sp^3$
bonds, resulting in a high speed of sound, make monocrystalline
diamond one of the best thermal conductors \cite{diamond-condmax}. An
unusually high thermal conductance should also be expected in carbon
nanotubes\cite{{Iijima91},{Dresselhaus96}}, which are held together
by even stronger $sp^2$ bonds. These systems, consisting of seamless
and atomically perfect graphitic cylinders few nanometers in
diameter, are self-supporting. The rigidity of these systems, combined
with virtual absence of atomic defects or coupling to soft phonon
modes of the embedding medium, should make isolated nanotubes very
good candidates for efficient thermal conductors. This conjecture has
been confirmed by experimental data that are consistent with a very
high thermal conductivity for nanotubes \cite{Zettl}.
 
In the following, we will present results of molecular dynamics
simulations using the Tersoff potential \cite{tersoff}, augmented by
Van der Waals interactions in graphite,  for the temperature
dependence of the thermal conductivity of nanotubes and other carbon
allotropes. We will show that isolated nanotubes are at least as good
heat conductors as high-purity diamond. Our comparison with graphitic
carbon shows that inter-layer coupling reduces thermal conductivity
of graphite within the basal plane by one order of magnitude with
respect to the nanotube value which lies close to that for a
hypothetical isolated graphene monolayer.
 
The thermal conductivity $\lambda$ of a solid along a particular
direction, taken here as the $z$ axis, is related to the heat flowing
down a long rod with a temperature gradient $dT/dz$ by
\begin{equation}
  \frac{1}{A}\frac{dQ}{dt} = - \lambda \frac{dT}{dz} \;,
\label{Eq1}
\end{equation}
where $dQ$ is the energy transmitted across the area $A$ in the time
interval $dt$. In solids where the phonon contribution to the heat
conductance dominates, $\lambda$ is proportional to $Cvl$, the
product of the heat capacity per unit volume $C$, the speed of sound
$v$, and the phonon mean free path $l$. The latter quantity is
limited by scattering from sample boundaries (related to grain
sizes), point defects, and by umklapp processes. In the experiment,
the strong dependence of the thermal conductivity $\lambda$ on $l$
translates into an unusual sensitivity to isotopic and other atomic
defects. This is best illustrated by the reported thermal
conductivity values in the basal plane of graphite
\cite{Landolt-Bornstein} which scatter by nearly two orders of
magnitude. As similar uncertainties may be associated with thermal
conductivity measurements in ``mats'' of nanotubes \cite{Zettl}, we
decided to determine this quantity using molecular dynamics
simulations.
 
The first approach used to calculate $\lambda$ was based on a direct
molecular dynamics simulation. Heat exchange with a periodic array of
hot and cold regions along the nanotube has been achieved by velocity
rescaling, following a method that had been successfully applied to
the thermal conductivity of glasses \cite{Jund}. Unlike glasses,
however, nanotubes exhibit an unusually high degree of long-range
order over hundreds of nanometers. The perturbations imposed by the
heat transfer reduce the effective phonon mean free path to below the
unit cell size. We found it hard to achieve convergence, since the
phonon mean free path in nanotubes is significantly larger than unit
cell sizes tractable in molecular dynamics simulations.
 
As an alternate approach to determine the thermal conductivity, we
used equilibrium molecular dynamics simulations
\cite{{schoen},{levesque}} based on the Green-Kubo expression that
relates this quantity to the integral over time $t$ of the heat flux
autocorrelation function by \cite{mcquarrie}
\begin{equation}
   \lambda=\frac{1}{3 V k_B T^2}
           \int_0^{\infty}<{\bf J}(t)\cdot{\bf J}(0)>dt \;.
\label{Eq2}
\end{equation}
Here, $k_B$ is the Boltzmann constant, $V$ is the volume, $T$ the
temperature of the sample, and the angled brackets denote an ensemble
average. The heat flux vector ${\bf J}(t)$ is defined by
\begin{eqnarray}
{\bf J}(t) &=& \frac{d}{dt} \sum_i {\bf r}_i {\Delta}e_i \\ \nonumber
           &=& \sum_i {\bf v}_i {\Delta}e_i
             - \sum_i\sum_{j({\ne}i)}{\bf r}_{ij}
                          ({\bf f}_{ij}\cdot{\bf v}_i) \;,
\label{Eq3}
\end{eqnarray}
where ${\Delta}e_i=e_i-<e>$ is the excess energy of atom $i$ with
respect to the average energy per atom $<e>$. ${\bf r}_i$ is the
position and ${\bf v}_i$ the velocity of atom $i$, and ${\bf
r}_{ij}={\bf r}_j-{\bf r}_i$. Assuming that the total potential
energy $U=\sum_i u_i$ can be expressed as a sum of binding energies
$u_i$ of individual atoms, then ${\bf f}_{ij}=-{\nabla}_i u_j$, where
${\nabla}_i$ is the gradient with respect to the position of atom
$i$.
 
In low-dimensional systems, such as nanotubes or graphene monolayers,
we infer the volume from the way how these systems pack in space 
(nanotubes form bundles and graphite a layered structure, both
with an inter-wall separation of ${\approx}3.4$~{\AA}) in
order to convert thermal conductance of a system to thermal
conductivity of a material.
 
Once ${\bf J}(t)$ is known, the thermal conductivity can be
calculated using Eq.~(\ref{Eq2}). We found, however, that these 
results depend
sensitively on the initial conditions of each simulation, thus
necessitating a large ensemble of simulations. This high
computational demand was further increased by the slow convergence of
the autocorrelation function, requiring long integration time periods.
 
\begin{figure}
    \centerline{
        {\raisebox{0.5\columnwidth}{\large\bf (a)}}
        \hspace*{-0.1\columnwidth}
        \epsfxsize=0.9\columnwidth
        \epsffile{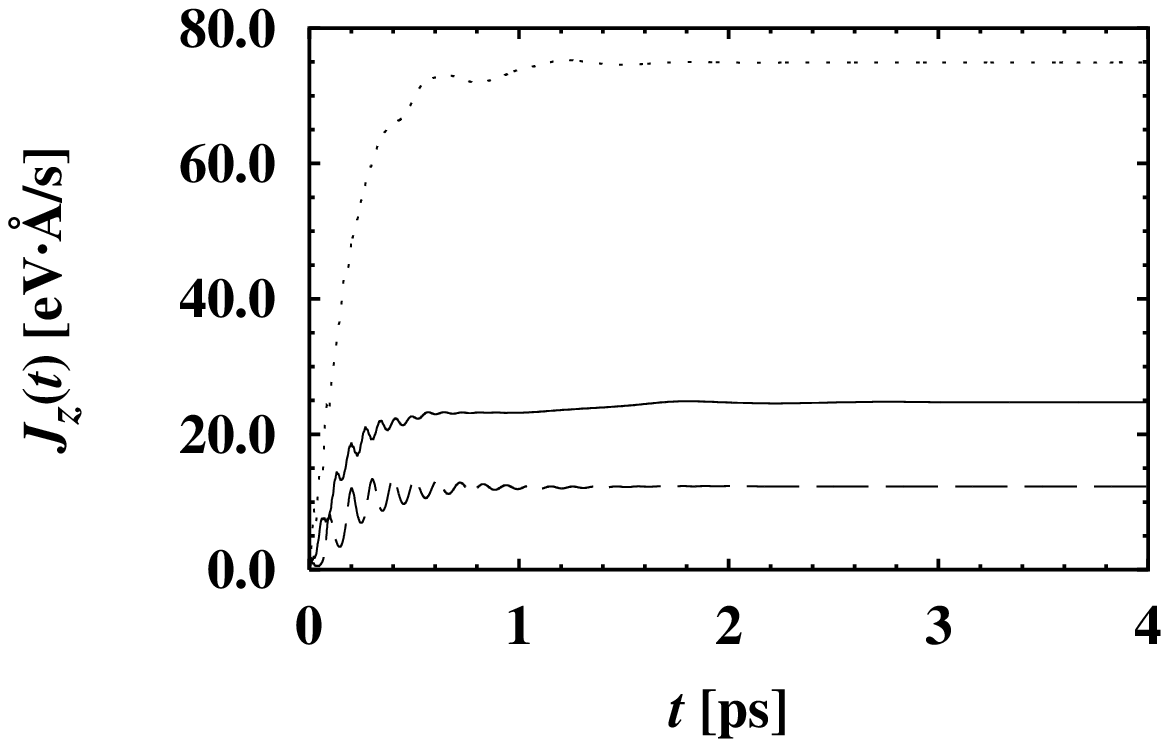}
    }
    \vspace*{-0.05\columnwidth}
    \centerline{
        {\raisebox{0.5\columnwidth}{\large\bf (b)}}
        \hspace*{-0.1\columnwidth}
        \epsfxsize=0.9\columnwidth
        \epsffile{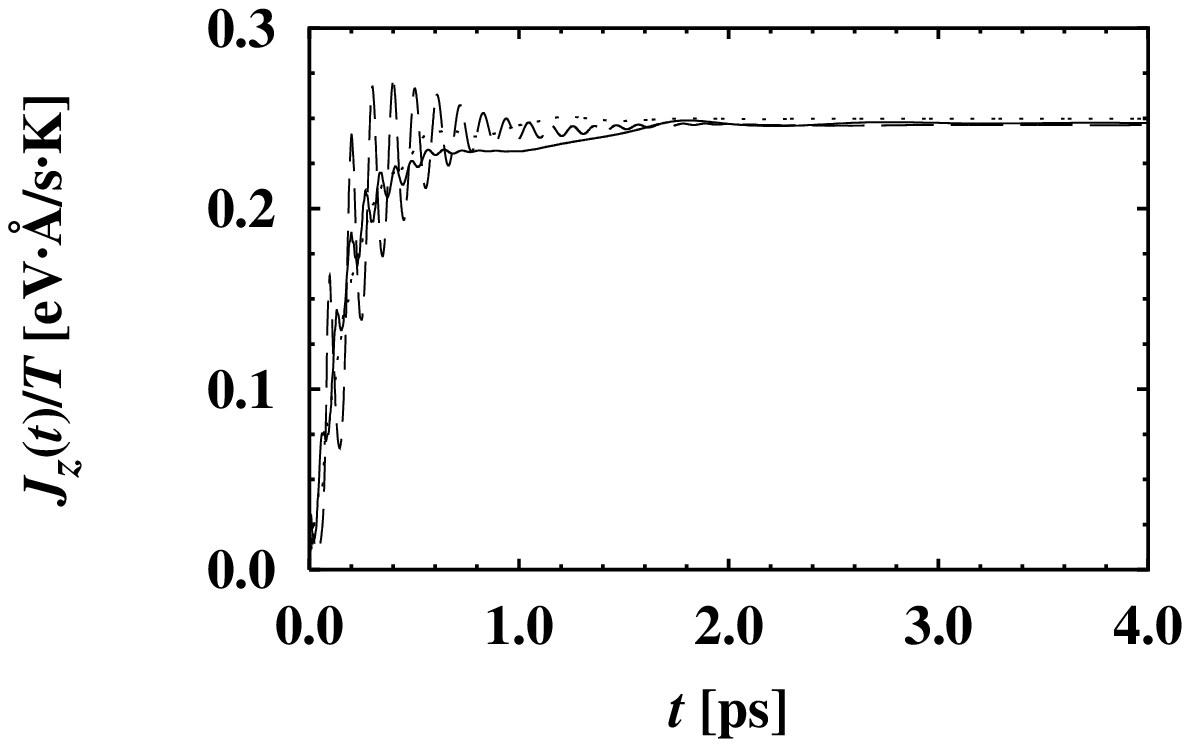}
    }
    \vspace*{-0.05\columnwidth}
    \centerline{
        {\raisebox{0.5\columnwidth}{\large\bf (c)}}
        \hspace*{-0.1\columnwidth}
        \epsfxsize=0.9\columnwidth
        \epsffile{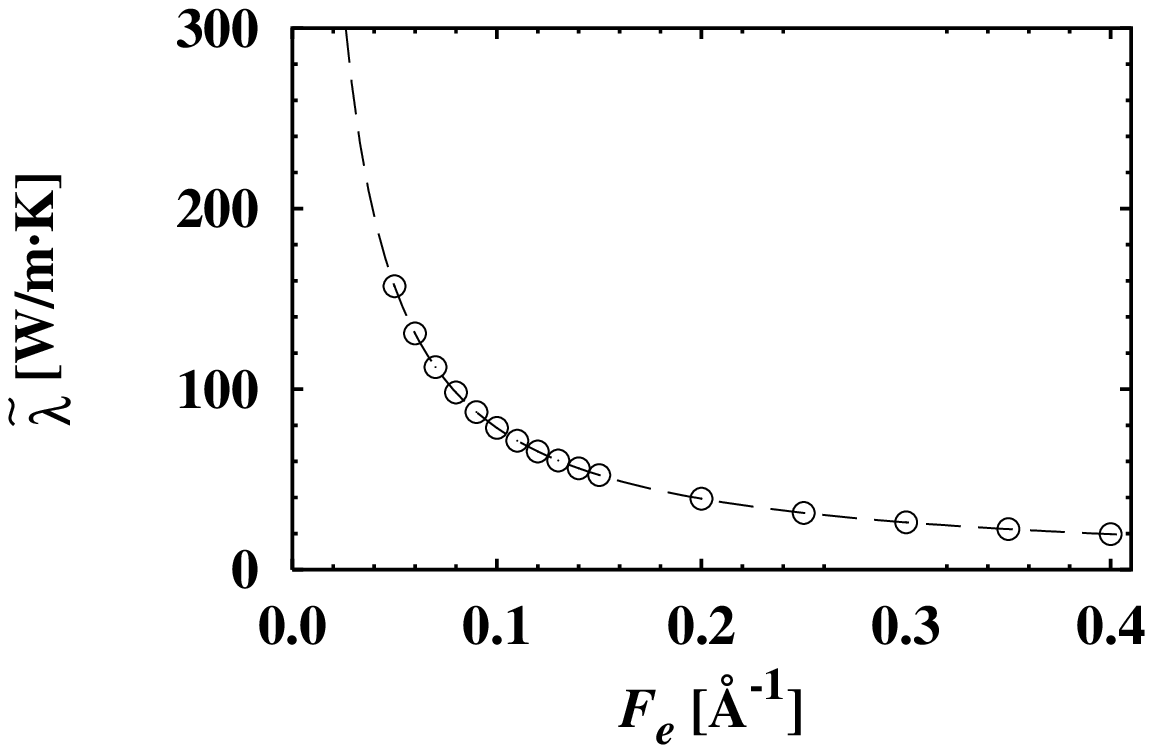}
    }
\caption{
(a) Time dependence of the axial heat flux $J_z(t)$ in a $(10,10)$
carbon nanotube. Results of nonequilibrium molecular dynamics
simulation at a fixed applied thermal force $F_e=0.2$~{\AA}$^{-1}$,
are shown at temperatures $T=50$~K (dashed line), $100$~K (solid
line), and $300$~K (dotted line).
(b) Time dependence of $J_z(t)/T$, a key quantity for the calculation
of the thermal conductivity, for $F_e=0.2$~{\AA}$^{-1}$ and the same
temperature values.
(c) Dependence of the heat transport on the applied heat force $F_e$
in the simulations for $T=100$~K. The dashed line represents an
analytical expression that is used to determine the thermal
conductivity $\lambda$ by extrapolating the simulation data points
$\tilde\lambda$ for $F_e{\rightarrow}0$.
}
\label{Fig1}
\end{figure}

These disadvantages have been shown to be strongly reduced in an
alternate approach \cite{maeda} that combines the Green-Kubo formula
with nonequilibrium thermodynamics \cite{{evans82},{evans94}} in a
computationally efficient manner \cite{Rapaport}. In this approach,
the thermal conductivity along the $z$ axis is given by
\begin{equation}
   \lambda = \lim_{{\bf F}_e{\rightarrow}0}
             \lim_{t{\rightarrow}\infty}
             \frac{<J_z({\bf F}_e,t)>}{F_e T V} \;,
\label{Eq4}
\end{equation}
where $T$ is the temperature of the sample, regulated by a
Nos\'e-Hoover thermostat \cite{Nose-Hoover}, and $V$ is the volume of
the sample. $J_z({\bf F}_e,t)$ is the $z$ component of the heat flux
vector for a particular time $t$. ${\bf F}_e$ is a small fictitious
``thermal force'' (with a dimension of inverse length) that is
applied to individual atoms. This fictitious force ${\bf F}_e$ 
and the Nos\'e-Hoover thermostat impose an additional force
${\Delta}{\bf F}_i$ on each atom $i$. This additional force modifies
the gradient of the potential energy and is given by
\begin{eqnarray}
{\Delta}{\bf F}_i &=& {\Delta}e_i {\bf F}_e -
             \sum_{j({\neq}i)} {\bf f}_{ij}
                                ({\bf r}_{ij}{\cdot}{\bf F}_e) \nonumber \\
        && + \frac{1}{N}\sum_j\sum_{k({\ne}j)} {\bf f}_{jk}
                                    ({\bf r}_{jk}{\cdot}{\bf F}_e)
           - \alpha {\bf p}_i \;.
\label{Eq5}
\end{eqnarray}
Here, $\alpha$ is the Nos\'e-Hoover thermostat multiplier acting on
the momentum ${\bf p}_i$ of atom $i$. $\alpha$ is calculated using
the time integral of the difference between the instantaneous kinetic
temperature $T$ of the system and the heat bath temperature $T_{eq}$,
from $\dot\alpha=(T-T_{eq})/Q$, where $Q$ is the thermal inertia. The
third term in Eq.~(\ref{Eq5}) guarantees that the net force acting on
the entire $N$-atom system vanishes.
 
In Fig.~\ref{Fig1} we present the results of our nonequilibrium
molecular dynamics simulations for the thermal conductance of an
isolated $(10,10)$ nanotube aligned along the $z$ axis. In our
calculation, we consider 400 atoms per unit cell, and use periodic
boundary conditions. Each molecular dynamics simulation run consists
of 50,000 time steps of $5.0{\times}10^{-16}$~s. Our results for the
time dependence of the heat current for the particular value
$F_e=0.2$~{\AA}$^{-1}$, shown in Fig.~\ref{Fig1}(a), suggest that
$J_z(t)$ converges within the first few picoseconds to its limiting
value for $t{\rightarrow}\infty$ in the temperatures range below
400~K. The same is true for the quantity $J_z(t)/T$, shown in
Fig.~\ref{Fig1}(b), the average of which is proportional to the
thermal conductivity $\lambda$ according to Eq.~(\ref{Eq4}). Our
molecular dynamics simulations have been performed for a total time
length of $25$~ps to represent well the long-time behavior.
 
\begin{figure}
    \centerline{
        \epsfxsize=0.9\columnwidth
        \epsffile{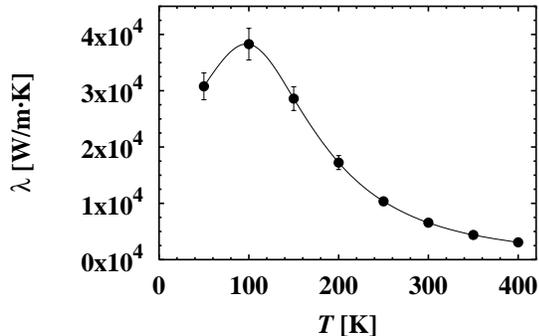}
    }
\caption{
Temperature dependence of the thermal conductivity $\lambda$ for a
$(10,10)$ carbon nanotube for temperatures below 400~K.
}
\label{Fig2}
\end{figure}

In Fig.~\ref{Fig1}(c) we show the dependence of the quantity
\begin{equation}
\tilde\lambda {\equiv}
\lim_{t\rightarrow\infty} \frac{<J_z({\bf F}_e,t)>}{F_e T V}
\label{Eq6}
\end{equation}
on $F_e$. We have found that direct calculations of $\tilde\lambda$
for very small thermal forces carry a substantial error, as they
require a division of two very small numbers in Eq.~(\ref{Eq6}). We
base our calculations of the thermal conductivity at each temperature
on 16 simulation runs, with $F_e$ values ranging from
$0.4-0.05$~{\AA}$^{-1}$. As shown in Fig.~\ref{Fig1}(c), data for
$\tilde\lambda$ can be extrapolated analytically for ${\bf
F}_e{\rightarrow}0$ to yield the thermal conductivity $\lambda$,
shown in Fig.~\ref{Fig2}.
 
Our results for the temperature dependence of the thermal conductivity
of an isolated $(10,10)$ carbon nanotube, shown in Fig.~\ref{Fig2},
reflect the fact that $\lambda$ is proportional to the heat capacity
$C$ and the phonon mean free path $l$. At low temperatures, $l$ is
nearly constant, and the temperature dependence of $\lambda$ follows
that of the specific heat. At high temperatures, where the specific
heat is constant, $\lambda$ decreases as the phonon mean free path
becomes smaller due to umklapp processes. Our calculations suggest
that at $T=100$~K, carbon nanotubes show an unusually high thermal
conductivity value of $37,000$~W/m$\cdot$K. This value lies very
close to the highest value observed in any solid,
$\lambda=41,000$~W/m$\cdot$K, that has been reported
\cite{diamond-condmax} for a 99.9\% pure $^{12}$C crystal at 104~K.
In spite of the decrease of $\lambda$ above 100~K, the room
temperature value of $6,600$~W/m$\cdot$K is still very high,
exceeding the reported thermal conductivity value of
$3,320$~W/m$\cdot$K for nearly isotopically pure diamond
\cite{diamond-condRT}.
 
\begin{figure}
    \centerline{
        \epsfxsize=0.9\columnwidth
        \epsffile{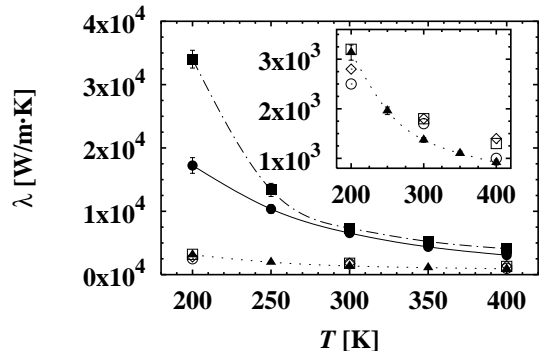}
    }
\caption{
Thermal conductivity $\lambda$ for a $(10,10)$ carbon nanotube (solid
line), in comparison to
a constrained graphite monolayer (dash-dotted line), and the basal
plane of $AA$ graphite (dotted line) at temperatures between $200$~K
and $400$~K. The inset reproduces the graphite data on an expanded
scale. The calculated values are compared to the experimental data of
Refs.\
\protect\onlinecite{Nihira} (open circles),
\protect\onlinecite{Holland} (open diamonds), and
\protect\onlinecite{Combarieu} (open squares) for graphite.
}
\label{Fig3}
\end{figure}

We found it useful to compare the thermal conductivity of a $(10,10)$
nanotube to that of an isolated graphene monolayer as well as bulk
graphite. For the graphene monolayer, we unrolled the 400-atom large
unit cell of the $(10,10)$ nanotube into a plane. The periodically
repeated unit cell used in the bulk graphite calculation contained
720 atoms, arranged in three layers.
The results of our calculations, presented in Fig.~\ref{Fig3},
suggest that an isolated nanotube shows a very similar thermal
transport behavior as a hypothetical isolated graphene monolayer, in
general agreement with available experimental data
\cite{{Nihira},{Holland},{Combarieu}}. Whereas even larger thermal
conductivity should be expected for a monolayer than for a nanotube,
we must consider that unlike the nanotube, a graphene monolayer is not
self-supporting in vacuum. For all carbon allotropes considered here,
we also find that the thermal conductivity decreases with increasing
temperature in the range depicted in Fig.~\ref{Fig3}.
 
Very interesting is the fact that once graphene layers are stacked in
graphite, the inter-layer interactions quench the thermal
conductivity of this system by nearly one order of magnitude. For the
latter case of crystalline graphite, we also found our calculated
thermal conductivity values to be confirmed by corresponding
observations in the basal plane of highest-purity synthetic graphite
\cite{{Nihira},{Holland},{Combarieu}} which are also reproduced in
the figure. We would like to note that experimental data suggest that
the thermal conductivity in the basal plane of graphite peaks near
100~K, similar to our nanotube results.
 
Based on the above described difference in the conductivity between a
graphene monolayer and graphite, we should expect a similar reduction
of the thermal conductivity when a nanotube is brought into contact
with other systems. This should occur when nanotubes form a bundle or
rope, become nested in multi-wall nanotubes, or interact with other
nanotubes in the ``nanotube mat'' of ``bucky-paper'' and could be
verified experimentally. Consistent with
our conjecture is the low value of $\lambda{\approx}0.7$~W/m$\cdot$K
reported for the bulk nanotube mat at room temperature \cite{Zettl}.
 
In summary, we combined results of equilibrium and non-equilibrium
molecular dynamics simulations with accurate carbon potentials to
determine the thermal conductivity $\lambda$ of carbon nanotubes and
its dependence on temperature. Our results suggest an unusually high
value ${\lambda}{\approx}6,600$~W/m$\cdot$K for an isolated $(10,10)$
nanotube at room temperature, comparable to the thermal conductivity
of a hypothetical isolated graphene monolayer or graphite. We believe
that these high values of $\lambda$ are associated with the large
phonon mean free paths in these systems. Our numerical data indicate
that in presence of inter-layer coupling in graphite and related
systems, the thermal conductivity is reduced significantly to fall
into the
experimentally observed value range.
 
This work was supported by the Office of Naval Research and DARPA
under Grant No. N00014-99-1-0252.
 

\end{document}